\documentclass[aip,jap,superscriptaddress,longbibliography,reprint,floatfix]{revtex4-1}

\usepackage{hyperref}
\usepackage{amsmath}
\usepackage{amssymb}
\usepackage{graphicx}

\usepackage{graphicx}

\usepackage{booktabs} 

\begin{document}

\title{X-ray detection of ultrashort spin current pulses in synthetic antiferromagnets}

\author{C.~Stamm}
\email[]{christian.stamm@fhnw.ch}
\affiliation{Department of Materials, ETH Zurich, 8093 Zurich, Switzerland}
\affiliation{Institute for Electric Power Systems, University of Applied Sciences and Arts Northwestern Switzerland, 5210 Windisch, Switzerland}

\author{C.~Murer}
\affiliation{Department of Materials, ETH Zurich, 8093 Zurich, Switzerland}

\author{M.\,S.~W{\"o}rnle}
\affiliation{Department of Materials, ETH Zurich, 8093 Zurich, Switzerland}
\affiliation{Department of Physics, ETH Zurich, 8093 Zurich, Switzerland}

\author{Y.~Acremann}
\affiliation{Laboratory for Solid State Physics, ETH Zurich, 8093 Zurich, Switzerland}

\author{R.~Gort}
\affiliation{Laboratory for Solid State Physics, ETH Zurich, 8093 Zurich, Switzerland}
\affiliation{European XFEL GmbH, Holzkoppel 4, 22869 Schenefeld, Germany}

\author{S.~D{\"a}ster}
\affiliation{Laboratory for Solid State Physics, ETH Zurich, 8093 Zurich, Switzerland}

\author{A.\,H.~Reid}
\affiliation{Linac Coherent Light Source, SLAC National Accelerator Laboratory, 2575 Sand Hill Road, Menlo Park, CA 94025, USA}

\author{D.\,J.~Higley}
\affiliation{Linac Coherent Light Source, SLAC National Accelerator Laboratory, 2575 Sand Hill Road, Menlo Park, CA 94025, USA}
\affiliation{Stanford Institute for Materials and Energy Sciences, SLAC National Accelerator Laboratory, 2575 Sand Hill Road, Menlo Park, CA 94025, USA}
\affiliation{Department of Applied Physics, Stanford University, Stanford, CA 94305, USA} 

\author{S.\,F.~Wandel}
\author{W.\,F.~Schlotter}
\affiliation{Linac Coherent Light Source, SLAC National Accelerator Laboratory, 2575 Sand Hill Road, Menlo Park, CA 94025, USA}

\author{P.~Gambardella}
\affiliation{Department of Materials, ETH Zurich, 8093 Zurich, Switzerland}


\begin{abstract}
We explore the ultrafast generation of spin currents in magnetic multilayer samples by applying fs laser pulses to one layer and measuring the magnetic response in the other layer by element-resolved x-ray spectroscopy. In Ni(5~nm)/Ru(2~nm)/Fe(4~nm), the Ni and Fe magnetization directions couple antiferromagnetically due to the RKKY interaction, but may be oriented parallel through an applied magnetic field. After exciting the top Ni layer with a fs laser pulse, we find that also the Fe layer underneath demagnetizes, with a $4.1 \pm 1.9$\% amplitude difference between parallel and antiparallel orientation of the Ni and Fe magnetizations. We attribute this difference to the influence of a spin current generated by the fs laser pulse that transfers angular momentum from the Ni into the Fe layer. Our results confirm that superdiffusive spin transport plays a role in determining the sub-ps demagnetization dynamics of synthetic antiferromagnetic layers, but also evidence large depolarization effects due to hot electron dynamics, which are independent of the relative alignment of the magnetization in Ni and Fe.

\end{abstract}

\maketitle

\section{Introduction}

The concept of spin currents, i.e.\ the flow of spin angular momentum in the absence of a net charge current, yields novel and challenging research topics, and empowers new technological opportunities in the field of spintronics \cite{BookSpinCurrent}.
In synthetic antiferromagnets, spin currents are a vital ingredient for mediating the dynamic coupling between the magnetic layers \cite{Duine2018}. 
Through the transfer of angular momentum from one magnetic layer to another, one can manipulate the magnetization, electrical resistance, and heat flow in both metallic and insulating systems.
A direct way to generate a spin current pulse is by exciting a magnetic thin film with an ultrafast laser pulse. The microscopic mechanism is exemplified by the generation of a superdiffusive spin current, which was previously introduced as a possible mechanism for ultrafast demagnetization \cite{Battiato2010}. 
In this model, laser excited non-equilibrium electrons move through the sample with different lifetimes for majority and minority electrons, leading to an effective transport of spins.
A remarkable consequence of this mechanism was demonstrated in Ni/Ru/Fe layered samples, in which two ferromagnetic films are magnetically decoupled by a Ru spacer layer. While the upper film (Ni) loses its magnetization upon laser excitation and thereby generates a spin current, the lower film (Fe) gains magnetic moment by absorbing the spin current, leading to an enhancement of its magnetization for a short time, before relaxation sets in. To detect this excess spins, VUV spectroscopy with fs pulses was used in an element-selective magneto-optic measurement of the magnetic state of the Ni and Fe films \cite{Rudolf2012,Turgut2013}.
By tuning the Ru thickness and the applied magnetic field, the magnetic coupling between the Ni and Fe films could be chosen to be either parallel or anti-parallel. Only for parallel alignment the magnetic moment in Fe was enhanced by laser pumping, whereas for antiparallel alignment both Ni and Fe lost magnetization upon pumping. This is considered to be the direct consequence of a superdiffusive spin current as modeled by theory \cite{Battiato2010,Battiato2012a}: whereas for parallel coupling the spins transferred from Ni  into Fe increase the Fe moment, in the antiparallel configuration the Ni spins lead to an additional decease of the Fe moment. 

A quantitative measurement of the involved spin current however is lacking so far, as the magneto-optic VUV measurements \cite{Rudolf2012,Turgut2013} hold no information on the absolute spin moments. Recent time and spin resolved photoemission studies indeed demonstrate the importance of measuring all possible spin-polarized states during fs laser excitation \cite{Eich2017,Gort2018}. Photoemission is however limited to probing only few atomic layers at the surface of the sample, and thus would provide information only on the top layer of this specific structure. For this reason we instead use x-ray absorption spectroscopy (XAS) which is capable of measuring the multilayer stack with elemental resolution and, through the x-ray magnetic circular dichroism (XMCD), provides information on the magnetic state of each layer.
Previous XMCD measurements with fs time resolution \cite{Stamm2007, Boeglin2010, Stamm2010, Radu2011} have been performed at the FemtoSpeX facility of the BESSY-II storage ring \cite{Holldack2014}. Investigating similar Ni/Ru/Fe samples we demonstrated the potential of the technique to investigate layer resolved dynamics in magnetic multilayers \cite{Eschenlohr2017}. However, due to the limited x-ray flux available at FemtoSpeX, the influence of the spin current could not be resolved, thus high intensity fs-pulsed x-ray sources such as today's free electron lasers (FEL) are needed for more precise measurements.

Here we used the intense and ultrashort x-ray pulses from the free electron laser LCLS at the SLAC National Accelerator Laboratory to quantitatively explore the magnetic response in the fs time domain following fs laser excitation of magnetic double layer samples. 
The key prerequisites for successful measurements with magnetic contrast are circularly polarized soft x-ray pulses. Their availability at LCLS was previously demonstrated by using the Delta undulator generating 98\% of circular polarization \cite{Lutman2016}, and high-quality XMCD spectra were recorded at the SXR instrument \cite{Higley2016,Higley2019}. Through the elemental resolved soft x-ray spectroscopy of the two magnetic layers we gain quantitative insight into the spin transport process on the fs time scale beyond the first demonstration of the effect \cite{Rudolf2012,Turgut2013}.

\section{Measurements}

\begin{figure}[tb]
	\includegraphics[width=8.5cm]{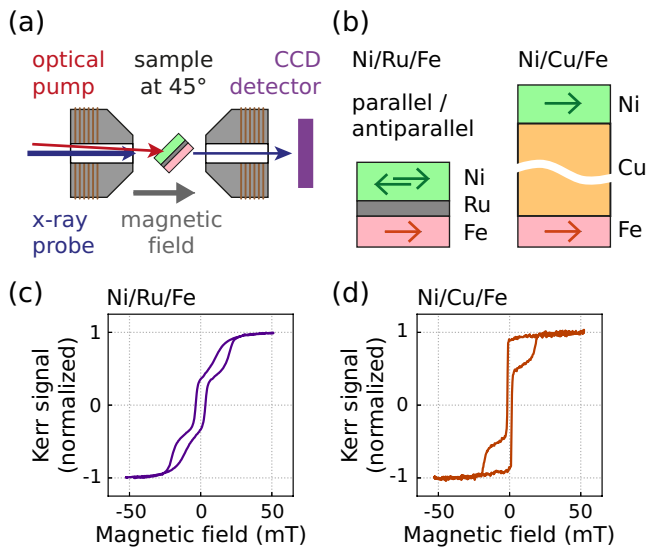} 
	\caption{(a) Measurement geometry:  optical pump beam, x-ray probe beam, and applied magnetic field are parallel at $45^{\circ}$ to the sample surface. The transmitted x-rays are measured on the CCD detector behind the sample. (b) Schematics of the multilayer samples, the Ni/Ru/Fe sample can have parallel and antiparallel magnetic alignment of the Ni and Fe layers. (c) Ni/Ru/Fe and (d) Ni/Cu/Fe magnetic hysteresis loops measured by longitudinal magneto-optic Kerr effect.}
	\label{setup}
\end{figure}

The x-ray transmission measurements have been performed at the LCLS SXR instrument \cite{Schlotter2012}, with the sample oriented at $45^\circ$ to the x-ray propagation direction as to measure the in-plane magnetization, see illustration in Fig.~\ref{setup}(a). A CCD detector is used to measure the integral intensity of the transmitted x-ray beam individually for each pulse, arriving with 120~Hz repetition rate. The pump pulse from the optical laser with 800~nm wavelength follows nearly the same path as the x-rays. A static magnetic field was applied during the measurements along the beam direction.

We investigated magnetic double layer samples of Ni and Fe, with Ru or Cu interlayer as drawn in Fig.~\ref{setup}(b). 
The Ni(5~nm)/Ru(2~nm)/Fe(4~nm) structure exhibits antiferromagnetic RKKY coupling between the Ni and Fe layers \cite{Parkin1991}. From the magnetic hysteresis loop in Fig.~\ref{setup}(c) we see that applying a sufficiently strong magnetic field (above $\approx 40$~mT) brings the magnetization directions of the Ni and Fe layers parallel, whereas in zero (or close to zero) magnetic field their magnetizations are oriented antiparallel.
The Ru layer thickness was optimized in 0.1~nm steps for best antiferromagnetic coupling between the Ni and Fe layers. The resulting 2~nm Ru is a bit thicker than the 1.5~nm used in  Ref.~[\onlinecite{Rudolf2012}], however both structures exhibit similar antiferromagnetic coupling with an exchange coupling field, given by the shift of the minor hysteresis loops, of $H_{ex}\approx 20$~mT.
In the Ni(5~nm)/Cu(30~nm)/Fe(4~nm) stack (Fig.~\ref{setup}(d)), the two magnetic layers are independent due to the relatively thick Cu layer, which is used to very efficiently block the pump pulse from the optical laser.
The samples were sputter deposited on top of a 200~nm thick Si$_3$N$_4$ square membranes of 250~$\mu$m edge length, on a 3~nm Ta seed layer, and were protected by a cap of 3~nm AlO$_x$. 
On the back of the Si$_3$N$_4$ membrane a 150~nm thick Al layer was deposited to help dissipate the heat from the pump laser pulses, and block the 800~nm radiation from reaching the CCD detector.
One Si sample chip carried an array of $4 \times 4$ membranes, so that we could quickly move to a fresh membrane on the same sample chip in between individual acquisition runs.

\subsection{ Time delay scan measurements on Ni/Ru/Fe}

\begin{figure}[tb]
	\includegraphics[width=6.5cm]{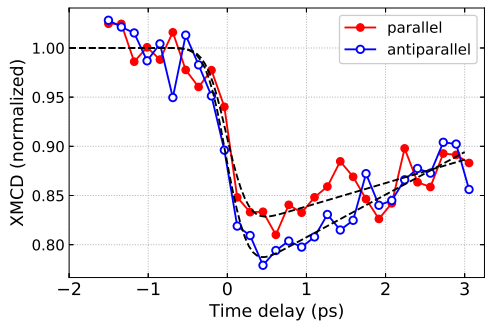} 
	\caption{Magnetic response of the Fe layer in Ni/Ru/Fe, measured as XMCD at the Fe $L_3$ edge, for parallel and antiparallel alignment of the Ni magnetization, at 26~mJ/cm$^2$ pump fluence. Dashed lines are fits.}
	\label{NiRuFe-timescan}
\end{figure}

We first discuss results from time scan measurements, for which we vary the pump probe time delay while keeping the probing x-ray photon energy fixed at the respective $L_3$ absorption line of Ni or Fe, and selecting circular polarization from the delta undulator. XMCD was calculated as the difference in absorption for the sample in positive and negative applied magnetic field.
In order to test the effect of a possible spin current from the (top) Ni layer into the (lower) Fe layer, we measured XMCD for the two relative orientations of their magnetizations, parallel and antiparallel. For the parallel state, we applied a field of 50~mT along the x-ray beam direction, or at 45$^\circ$ to the surface normal. To reach the antiparallel configuration, we first saturated the sample in 50~mT and then measured in a reduced field of $<10$~mT, which was used to define the preferential magnetization direction for the relaxation process following the pump laser pulse.
We show in Fig.~\ref{NiRuFe-timescan} the magnetic response of the Fe layer after excitation of the Ni/Ru/Fe sample. The two time traces correspond to the two relative orientations of the Ni and Fe magnetization directions. 
Within a time interval of $\approx 2$~ps after pumping the stack, the Fe XMCD curves deviate from each other, with a stronger magnetization reduction of the Fe layer in the antiparallel case. 
In this configuration, electrons that are directly excited in the Ni layer and move into the Fe film would contribute to the loss of Fe magnetization. For parallel orientation of the Ni and Fe spins, the transferred Ni electrons on the other hand would add to the magnetic moment of the Fe host's electrons, leading to weaker demagnetization of the Fe layer. In previous experiments, this even led to a transient increase of magnetic signal beyond the equilibrium value \cite{Rudolf2012}.
We may conclude that the observed difference between the two orientations' XMCD response is due to a spin current pulse launched in Ni by the optical pump pulse, which transfers spins into the lower Fe layer following the concept of superdiffusive spin transport \cite{Battiato2010}. 
The effect is rather short-lived, being visible only within $\approx 2$~ps after excitation. This time corresponds to the expected relaxation of the hot electrons and the cascades of secondary electrons that were excited in the process.

\subsection{Energy scan measurements on Ni/Ru/Fe}

\begin{figure}[htb]
	\includegraphics[width=6.5cm]{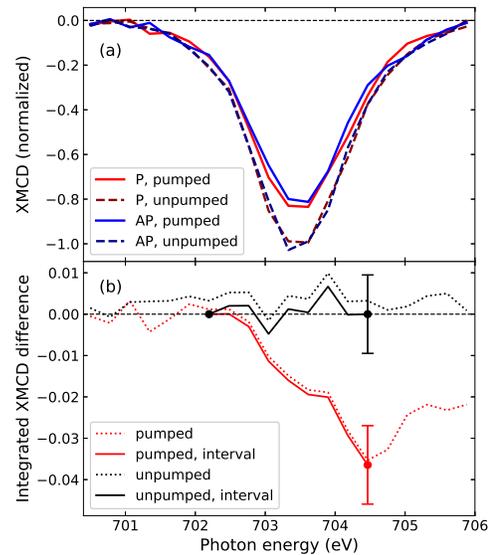} 
	\caption{(a) Ni/Ru/Fe XMCD spectra at the Fe $L_3$ absorption edge, measured 0.7~ps after pumping with 26~mJ/cm$^2$. (b) Integrated difference (P-AP) of the XMCD, for pumped and unpumped samples, scaled to the XMCD value of the unpumped sample. The full lines depict an integration start at the $L_3$ shoulder, whereas integration over the whole energy range is shown as dotted lines.}
	\label{NiRuFe-energyscan}
\end{figure}

For a more detailed look at the electronic state of the excited sample we performed Fe $L_3$ energy scans selecting a pump probe time delay of $\approx 1$~ps, i.e.\ shortly after laser pumping. Applying different magnetic fields to achieve the parallel and antiparallel configuration led to slightly different XMCD amplitudes. We thus alternated between measuring with and without optical pump pulse,
and normalize the XMCD spectra to the respective measurements without laser pump pulse.
We plot in Fig.~\ref{NiRuFe-energyscan}(a) the resulting XMCD for both relative magnetization directions, with and without applied pump pulse. 
The unpumped XMCD curves coincide within the measurement error, as expected. In contrast, there are small but significant differences between the pumped curves:
in the parallel state, the magnetization is better conserved as in the antiparallel state. 
This is qualitatively the same result as we found in the time scans in Fig.~\ref{NiRuFe-timescan}.  
We now determine the magnetic response by integrating the XMCD across the $L_3$ edge, and  plot in Fig.~\ref{NiRuFe-energyscan}(b) the difference of the parallel-antiparallel states. This graph best illustrates the effect of the spin current pulse on the magnetization of the Fe layer, and gives an estimate of the experimental uncertainty through the indicated error bars.

\begin{figure}[tb!]
	\includegraphics[width=6.5cm]{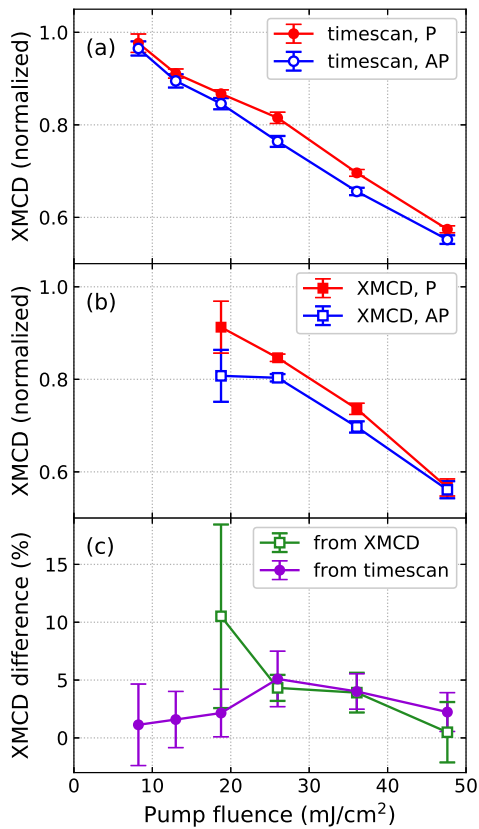} 
	\caption{XMCD of the Fe layer in Ni/Ru/Fe vs.\ pump fluence.
	(a) evaluated from timescans at photon energy set to $L_3$, and (b) from integration of XMCD spectra around $L_3$, for parallel (P) and antiparallel (AP) alignment of the Ni and Fe magnetization directions as indicated. (c) XMCD difference between P and AP configuration.}
	\label{fluence}
\end{figure}

In order to find the strongest effect of the spin current on the Fe layer underneath, we repeated the time and energy scans at different values for the pump fluences, and show the resulting XMCD values at $\approx$1~ps after excitation in Fig.~\ref{fluence}(a,b) for the time scan and energy scan data, respectively. The effect of the spin current, i.e.\ the XMCD difference between parallel and antiparallel, is plotted in Fig.~\ref{fluence}(c). It is highest for the central fluences between 26-36~mJ/cm$^2$, where it amounts to $4.1 \pm 1.9$\%.
We argue that this influence of the Ni magnetization direction on the size of the demagnetization in the Fe layer is caused by a spin current flowing from Ni into Fe.
Whereas for smaller fluences the effect is too small to be clearly detected, at the highest pump fluence a significant amount of direct excitation in the Fe layer apparently suppresses the effect.
It remains to discuss the overall magnetization drop in Fe, which is indeed the predominant effect in the Fe layer. A calculation of the absorption of the exciting laser pulse using a transfer matrix approach shows that the Fe layer absorbs 68\% of the energy absorbed in the Ni layer (see supplementary material). This is caused by the relatively thin 2~nm Ru interlayer, which can only block part of the laser beam from reaching the Fe layer.
As a consequence, the Fe layer experiences a sizable direct optical excitation, which leads to the observed overall demagnetization.

\begin{figure}[tb]
	\includegraphics[width=6.5cm]{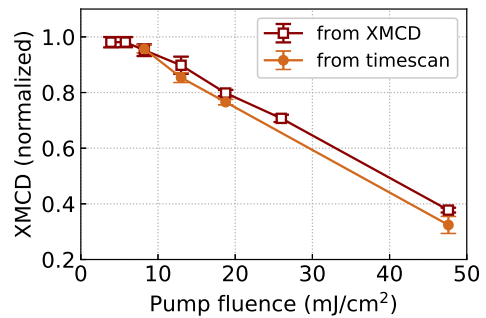} 
	\caption{Normalized XMCD of the Fe layer in Ni/Cu/Fe vs.\ pump fluence, evaluated from integration of XMCD spectra around the Fe $L_3$ edge, and from timescans at photon energy set to Fe $L_3$.}
	\label{fluence-NiCuFe}
\end{figure}

\subsection{Measurements on Ni/Cu/Fe}

We thus additionally investigated a multilayer consisting of Ni(5~nm)/Cu(30~nm)/Fe(4~nm), in which the thick Cu interlayer was introduced to block the optical pump laser. In fact, calculating the absorbed intensity in each layer with the transfer matrix method, we find that the Fe layer absorbs only 1\% of the intensity that the Ni layer absorbs, see supplementary material. For the Ni/Cu/Fe sample, we show in Fig.~\ref{fluence-NiCuFe} the resulting excitation on the Fe magnetization as function of the pump fluence. For this sample, the Ni and Fe magnetization directions were always aligned parallel, thus we would expect an enhancement of the Fe magnetization, if the excitation would be only by superdiffusive spin transport.
We observe however a monotonic decrease of the Fe XMCD with increasing fluence, consistently seen in both types of measurement, XMCD spectra at fixed pump-probe time delay and timescans at fixed photon energy. Only at low pump fluences around $\approx$5~mJ/cm$^2$ we may see a plateau, or threshold behavior, but our signal quality is not sufficient to indicate any enhancement effect. 
Hence, for the Ni/Cu/Fe trilayer the spin current was not strong enough to lead to a sizable deviation from a pure magnetization drop.

\section{Discussion}

Comparing the results from the two samples, we notice a much stronger Fe XMCD loss in Ni/Cu/Fe than in the Ni/Ru/Fe sample. The calculated absorption value for the top Ni layer may reveal the reason: the direct optical excitation of the Ni film in Ni/Cu/Fe is almost 10$\times$ higher than in Ni/Ru/Fe, as a consequence of the semi-transparent Ni/Ru/Fe stack in conjunction with a highly reflective Al layer on the back of the membrane. The higher reflectivity, i.e.\ lower absorption, compared to the Ni/Cu/Fe sample therefore leads to a considerably higher number of hot electrons that are generated in the latter, for the same incoming pump fluence.
We have seen before in our measurements on Ni/Ru/Fe that the spin dependent response in Fe is quenched at very high fluences, being overwhelmed presumably by direct laser excitation. In Ni/Cu/Fe we efficiently block the pump laser from reaching the Fe layer, however we cannot block hot electrons from moving between the layers. Together with the produced cascades of secondary electrons, the hot electrons transport spin but also energy towards the Fe layer. It was shown before that excited electrons can have greatly enhanced spin-flip scattering rates \cite{Koopmans2005}, a mechanism that would lead to ultrafast demagnetization of the lower Fe layer also in the Ni/Cu/Fe sample.
It looks as if these high fluences dominate the superdiffusive spin transport mechanism, or alternatively that non-equilibrium electrons may be affected unusually strong by depolarizing effects in the Cu interlayer, a material that is otherwise known to have a much longer spin diffusion length of 400-700~nm at room temperature \cite{Kimura2008,Cai2019}.
Besides the ultrafast excitation, the absorbed energy also leads to an increased average temperature of the sample. Examining XMCD spectra of the sample in between two pump pulses, we noticed a small decrease of magnetic moment when increasing the pump fluence from 19 to 48~mJ/cm$^2$, from which we estimate a rise of the sample's base temperature of about 40 up to 100~K. The presumably higher electron scattering rates at elevated temperature may be modeled by a shortened spin diffusion length \cite{Jedema2001,Yakata2006}, and can therefore lead to a less effective spin transport through the sample.

Finally we discuss quantitatively the observed spin current. The XMCD time scans at the Ni $L_3$ edge performed at 23 and 36~mJ/cm$^2$ (see supplementary material) allow us to interpolate a 55\% drop of the Ni magnetization for the fluence of 26~mJ/cm$^2$ used for the relevant Fe scan. With a Ni magnetization of 485~kA/m, this amounts to a loss of 267~kA/m, available for the spin current originating from the Ni film. 
Assuming a lossless spin transport we find, after correcting by 5/4 for the Ni and Fe layer thicknesses, a magnetization of 333~kA/m ready to flow into Fe. This is as much as 20\% of the Fe magnetization, assuming the full Fe magnetization of 1700~kA/m. 
Our measurements report an effect on the Fe magnetization of 4\%, i.e.\ $\pm 2$\%  deviation from the central value, 
corresponding to $\pm 34$~kA/m. Therefore, we conclude that only 10\% of the total Ni magnetization is transferred to Fe as a spin current. At least two effects will attenuate the spin current from Ni to Fe. One is the direct spin relaxation within Ni, as found for laser-induced demagnetization for single layer Ni films \cite{Beaurepaire1996,Koopmans2005,Stamm2007,Roth2012}. The other is spin relaxation during the transport of the spin current in the bulk or at the interfaces.
However, the pure difference in the Ru layer thickness, which was 2~nm compared to 1.5~nm in Ref.~[\onlinecite{Rudolf2012}], is much smaller than the 2-4~nm estimated spin diffusion length for Ru \cite{Yakata2006}, and thus cannot explain by itself the observed difference in spin current size.

\section{Conclusion}

Here we presented time and layer-resolved XMCD measurements on trilayer samples consisting of two magnetic layers, Ni and Fe, separated by a non-magnetic Ru or Cu interlayer. Whereas the top magnetic layer was excited by an optical fs laser pulse and subsequently exhibited ultrafast demagnetization, we measured the response of the lower magnetic layer which reacted with a less pronounced magnetization drop.
For a 2~nm Ru interlayer, the Ni and Fe layer's magnetization direction were oriented antiparallel, making the structure a synthetic antiferromagnet. Application of a magnetic field can align the Ni and Fe magnetization parallel, and we found that the amount of Fe demagnetization depends on the relative magnetization orientation of the two layers. 
We attribute this to the influence of spin currents transporting angular momentum from Ni into Fe, and observed this difference to be $4.1 \pm 1.9$\% of the initial Fe magnetization. This is much smaller than the 25-30\% reported in previous measurements that used VUV pulses as a probe \cite{Rudolf2012}. 
This is surprising, as we used similar samples and the excitation was nearly the same in both types of measurements.
We are left to speculate if the measurement method, probing through either VUV or soft x-rays, or the exact sample preparation has enough influence to cause this discrepancy.
Furthermore, our results suggest that superdiffusive spin transport plays a visible role in laser-excited multilayers, but depending on the exact sample geometry and excitation parameters, other effects caused by hot electron dynamics may be dominant.
Future experiments may shed light on the influence of electron scattering and depolarization properties in highly non-equilibrium environments, in order to understand how to optimize the spin transport in synthetic antiferromagnets.

\section*{Supplementary Material}

The following contents are included in the supplementary material: \\
Signal correlation and detector linearity \\
Pump probe timing: 60~Hz jitter correction \\
Additional time scans on Ni/Ru/Fe \\
Additional spectra on Ni/Ru/Fe \\
fs laser induced change in XAS \\
Additional data on Ni/Cu/Fe \\
Modeling of the laser absorption in multilayer samples

\begin{acknowledgments}
We thank A.\ Lutman and the SLAC operations team for setting up the circularly polarized x-ray beam, S.\ Zohar for help in the data processing, and L.\ Le Guyader for discussions on time jitter correction methods.
Use of the Linac Coherent Light Source (LCLS), SLAC National Accelerator Laboratory, is supported by the U.S. Department of Energy, Office of Science, Office of Basic Energy Sciences under Contract No. DE-AC02-76SF00515.
We acknowledge financial support from the Swiss National Science Foundation through grants No.\ 200021-153404 and 200020-172775.
\end{acknowledgments}

\section*{Data Availability Statement}
Raw data were generated and are hosted at the SLAC National Accelerator Laboratory.  Derived data supporting the findings of this study are available from the corresponding author upon reasonable request.


%


\clearpage
\onecolumngrid{
	\center
	\bf{
		{\Large Supplemental Material} \\
		\bigskip \bigskip
	}
}

\setcounter{equation}{0}
\setcounter{figure}{0}
\setcounter{table}{0}
\setcounter{page}{1}
\setcounter{section}{0}
\makeatletter
\renewcommand{\thesection}{SM \arabic{section}}
\renewcommand{\theequation}{S\arabic{equation}}
\renewcommand{\thefigure}{S\arabic{figure}}
\renewcommand{\bibnumfmt}[1]{[S#1]}
\renewcommand{\citenumfont}[1]{S#1}
\renewcommand{\thetable}{S\arabic{table}}	

\renewcommand{\arraystretch}{1.1} 

\newcommand\scalemath[2]{\scalebox{#1}{\mbox{\ensuremath{\displaystyle #2}}}}
\newcommand\mycom[2]{\genfrac{}{}{0pt}{}{#1}{#2}}


\section{Signal correlation and detector linearity}

\begin{figure}[h]
	\includegraphics[width=12cm]{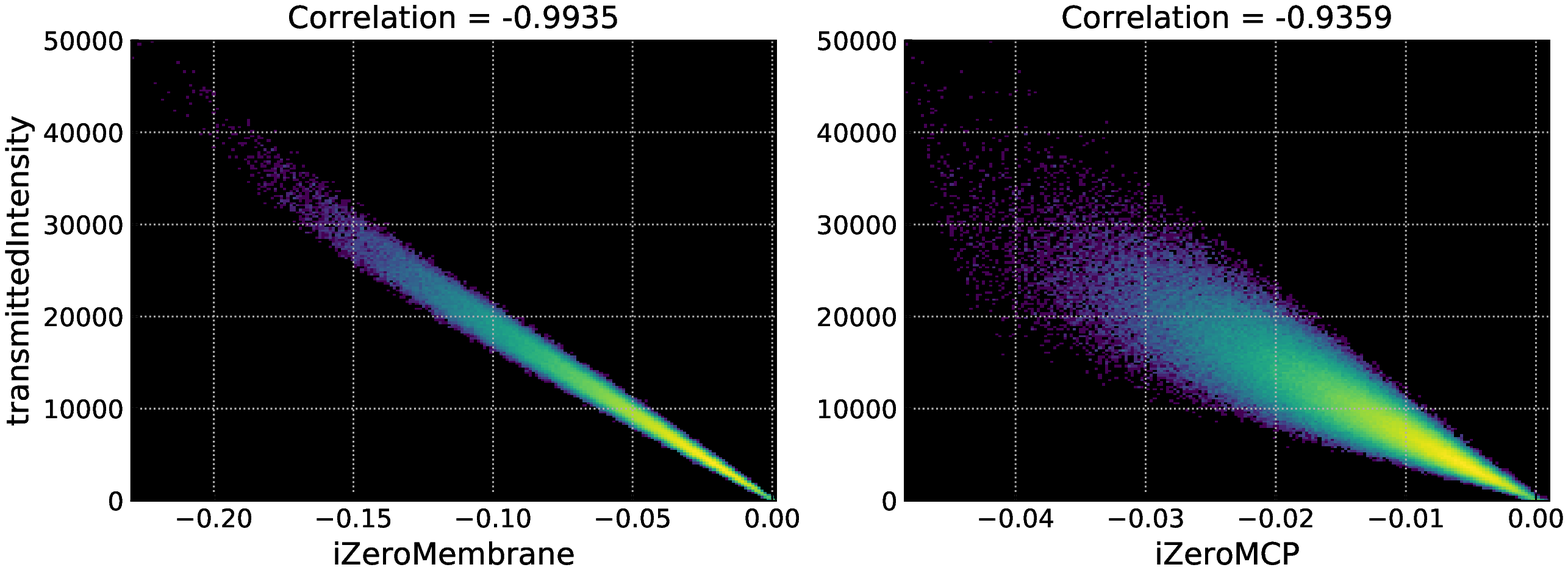}
	\caption{Correlation of the transmitted intensity measured on a CCD, with the incoming intensity recorded either at a Si membrane placed in the x-ray beam, or on a MCP that detects fluorescence and secondary electrons from an x-ray mirror.}
	\label{correlation}
\end{figure}

In order to measure the x-ray absorption, we recorded the transmitted x-ray intensity on a CCD detector by integration over all pixels, on a shot to shot basis with 120~Hz repetition rate. These values are divided by the incoming intensity, which is measured using two different methods. 
A fluorescence intensity monitor, operated in its MCP mode, detected both fluorescence and electrons from an x-ray mirror \cite{Heimann2019}. In addition, an MCP detector looking at the fluorescence from a Silicon membrane was used.
Detector linearity is ensured by plotting the correlation of the two measured signals in Fig.~\ref{correlation}. For the normalization of our measurements, we selected the detector at the Silicon membrane, as it had a higher correlation with the transmitted signal.

\section{Pump probe timing: 60~H\lowercase{z} jitter correction \label{jitter-corr}}

\begin{figure}[ht]
	\includegraphics[scale=0.5]{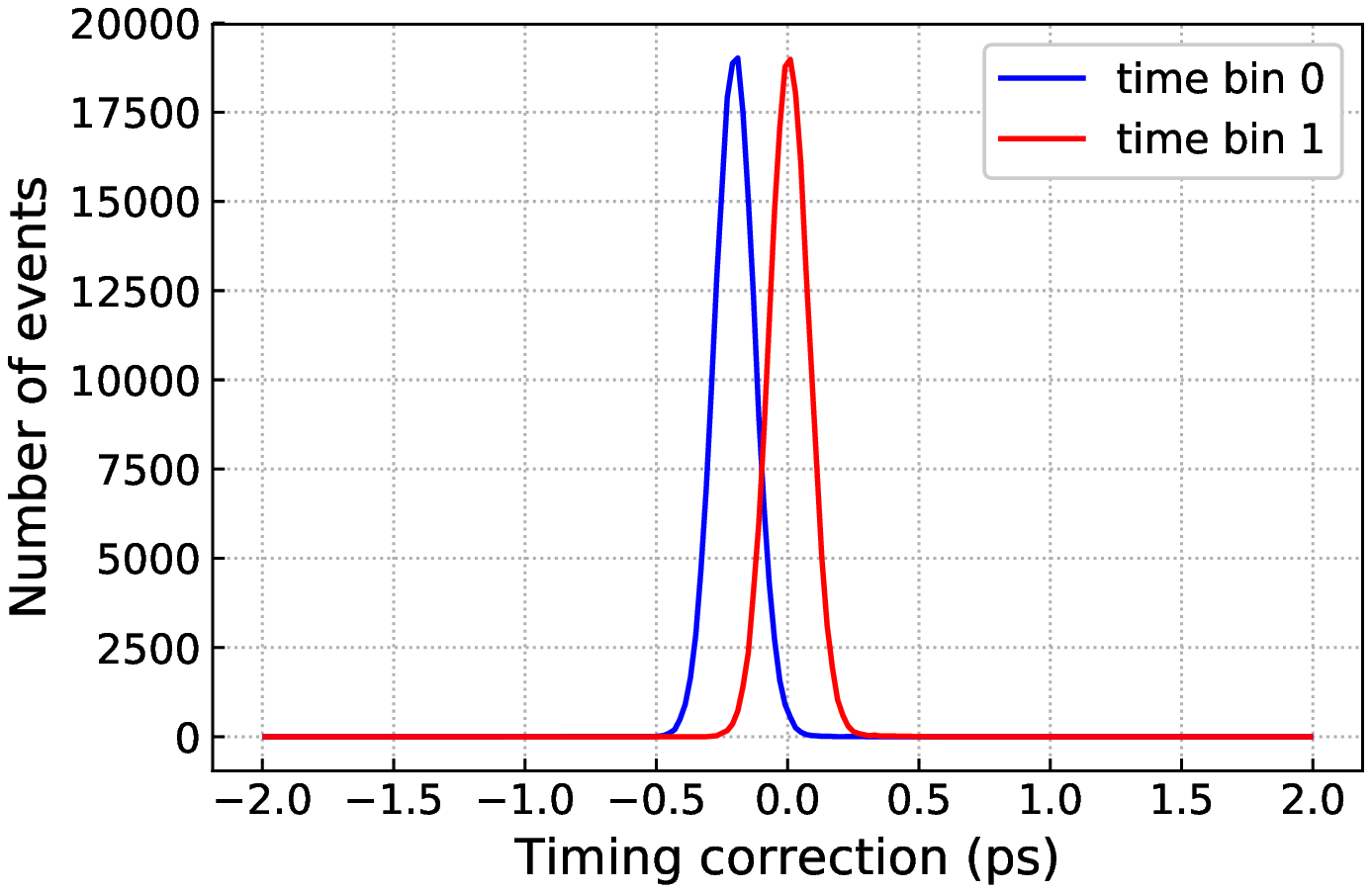}
	\includegraphics[scale=0.5]{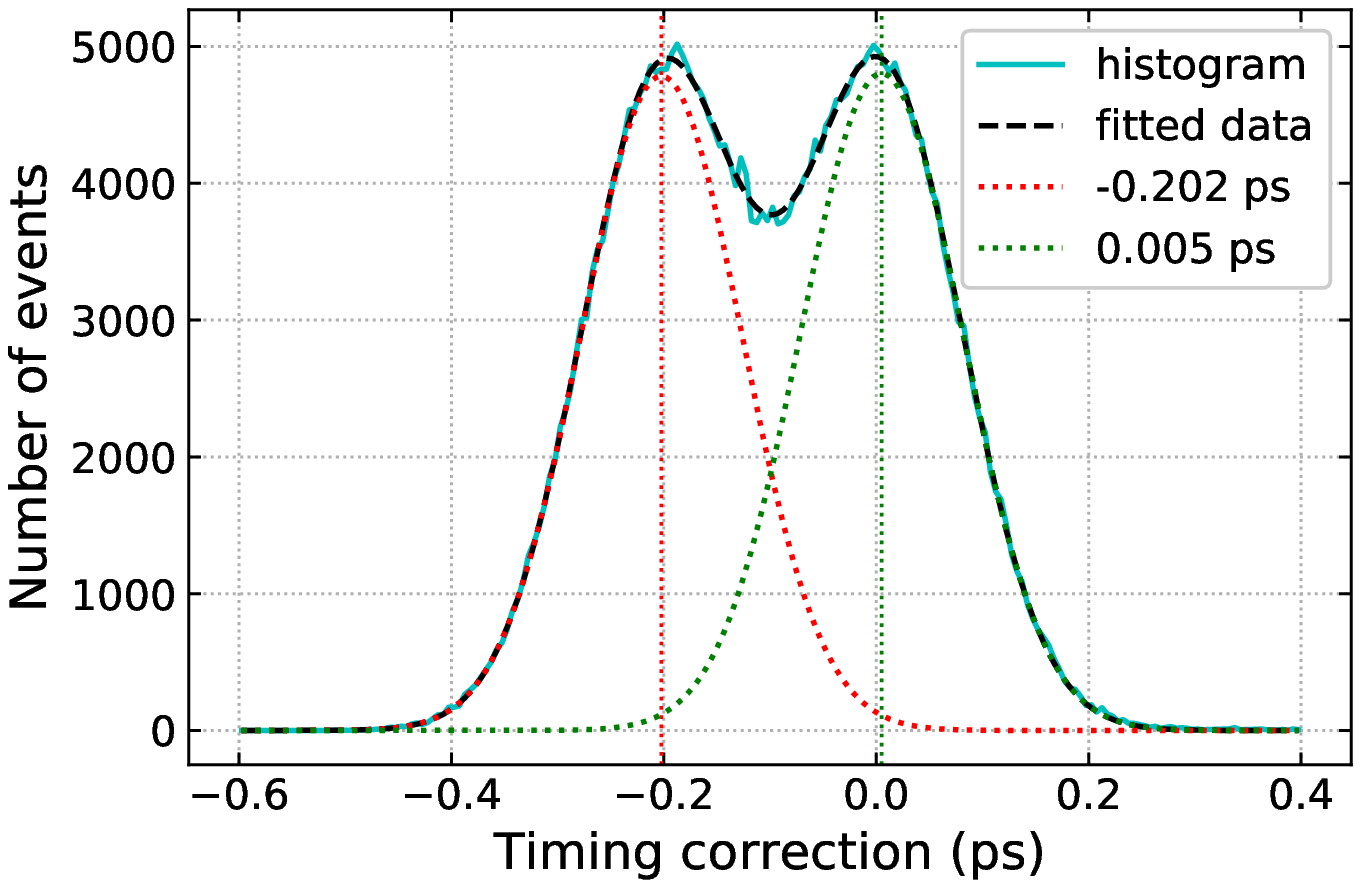}
	\caption{Phase cavity diagnostic of the arrival time correction. Left: histogram of the phase cavity arrival time correction, separated into two bins according to odd/even pulse number. Right: fitting a double Gaussian pulse reveals the time shift between the bins to be 0.207~ps.}
	\label{phase_cavity}
\end{figure}

Even with x-ray and laser pulses of sub-100 fs duration, the time resolution of a laser pump -- x-ray probe experiment may be worse due to arrival time jitter of the individual pulses. A known fact is that the LCLS electron bunch, and with it the generated x-ray pulse, has a timing jitter alternating with 60~Hz, for 120~Hz pulse repetition. Several options for handling the timing jitter have been tested in the data analysis, and we selected the second one for the presented graphs for stability reasons, and confirmed its credibility by comparison with the third option.
\begin{itemize}
	\item 
	Use original data without any time jitter correction: useful for quick online visualization of results. Depending on the jitter amplitude, XMCD decay times may increase to $> 500$~fs.
	\item
	Divide measurements into two 60~Hz bins, in order to separate the two alternating states with different x-ray arrival times. After that, perform individual fits of the XMCD data. For each bin, subtract their time-zero before combining to a common data set.
	\item
	Apply same division of the data as above. Use the phase cavity information that monitors the electron bunch arrival (see Fig.~\ref{phase_cavity}) to obtain the average time shift between the 60~Hz bins, then shift time-zero of one bin accordingly before combining to a common data set.
	This procedure gives similar results as the XMCD fitting method above, and does not need a measured dynamic signal from the sample.
	\item
	Use phase cavity data for shot-to-shot based correction. This procedure may have more noise, depending on data quality. Also long-term drifts of e.g. the pump laser line are not accounted for (see next two points for their inclusion).
	\item
	Apply a custom sorting algorithm based on the best correlation in the phase cavity corrected data from above, after averaging a few minutes of data
	\cite{LeGuyaderNote} outlined in \cite{Woody1967}, improved in \cite{Cabasson2009} and combined with a maximum-likelihood estimation \cite{Kim2016}. 
	If strong time-dependent changes are absent from the data, this method cannot be applied.
	\item
	Use time tool data, filtered by 100-1000 events to extract long-term drifts from the high noise level present in the timing data while operating with the delta undulator. For shot-to-shot correction, the phase cavity data is used. While this procedure has the best potential for precise jitter correction, unfortunately only very few data sets recorded with the circular polarization from the delta undulator have valid time tool data, due to the one order of magnitude lower x-ray intensity.
	
\end{itemize}

\section{Additional time scans on N\lowercase{i}/R\lowercase{u}/F\lowercase{e}}

\begin{figure}[ht]
	\centering
	\begin{tabular}{@{}cc@{}}
		\includegraphics[scale=0.5]{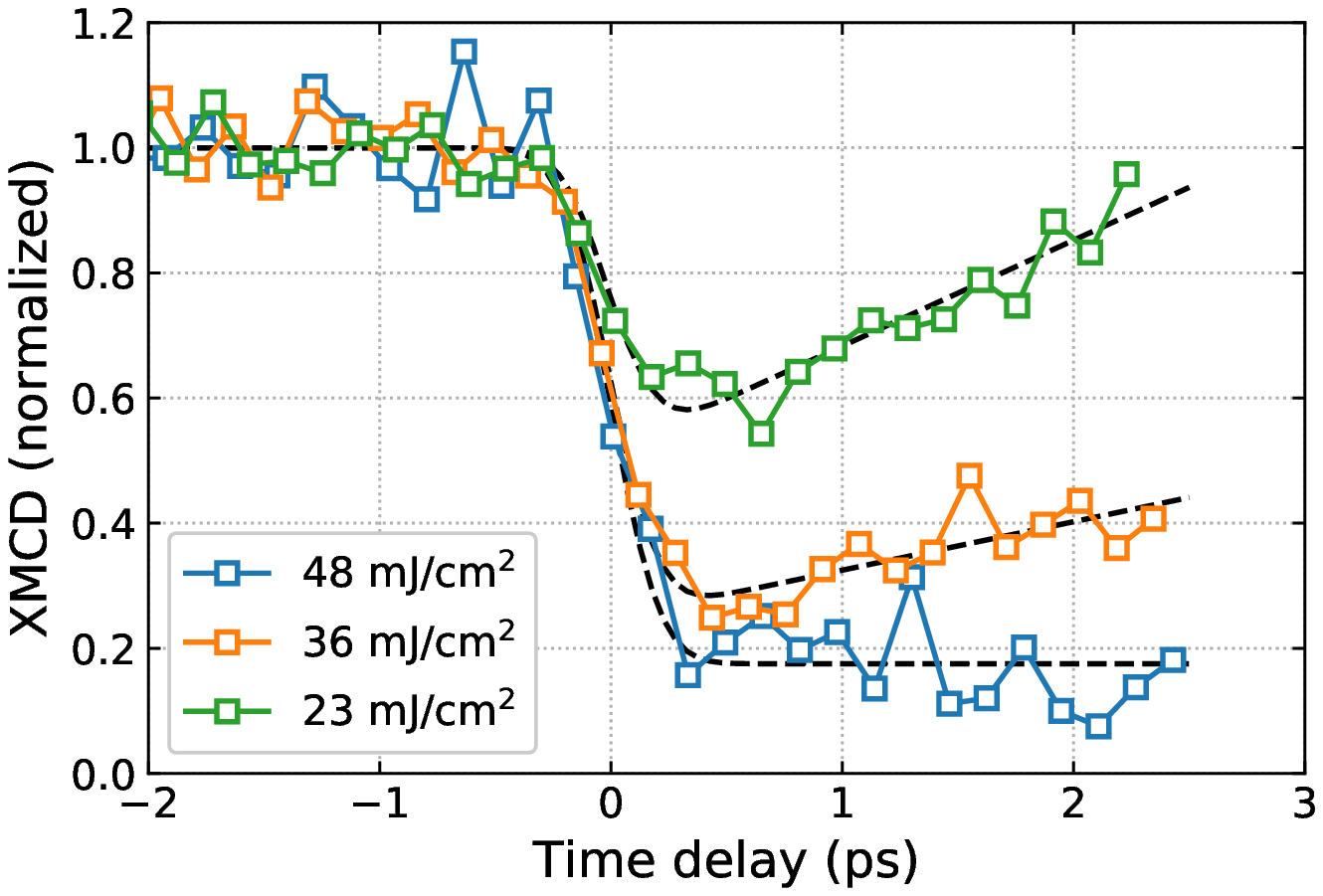} &
		\includegraphics[scale=0.5]{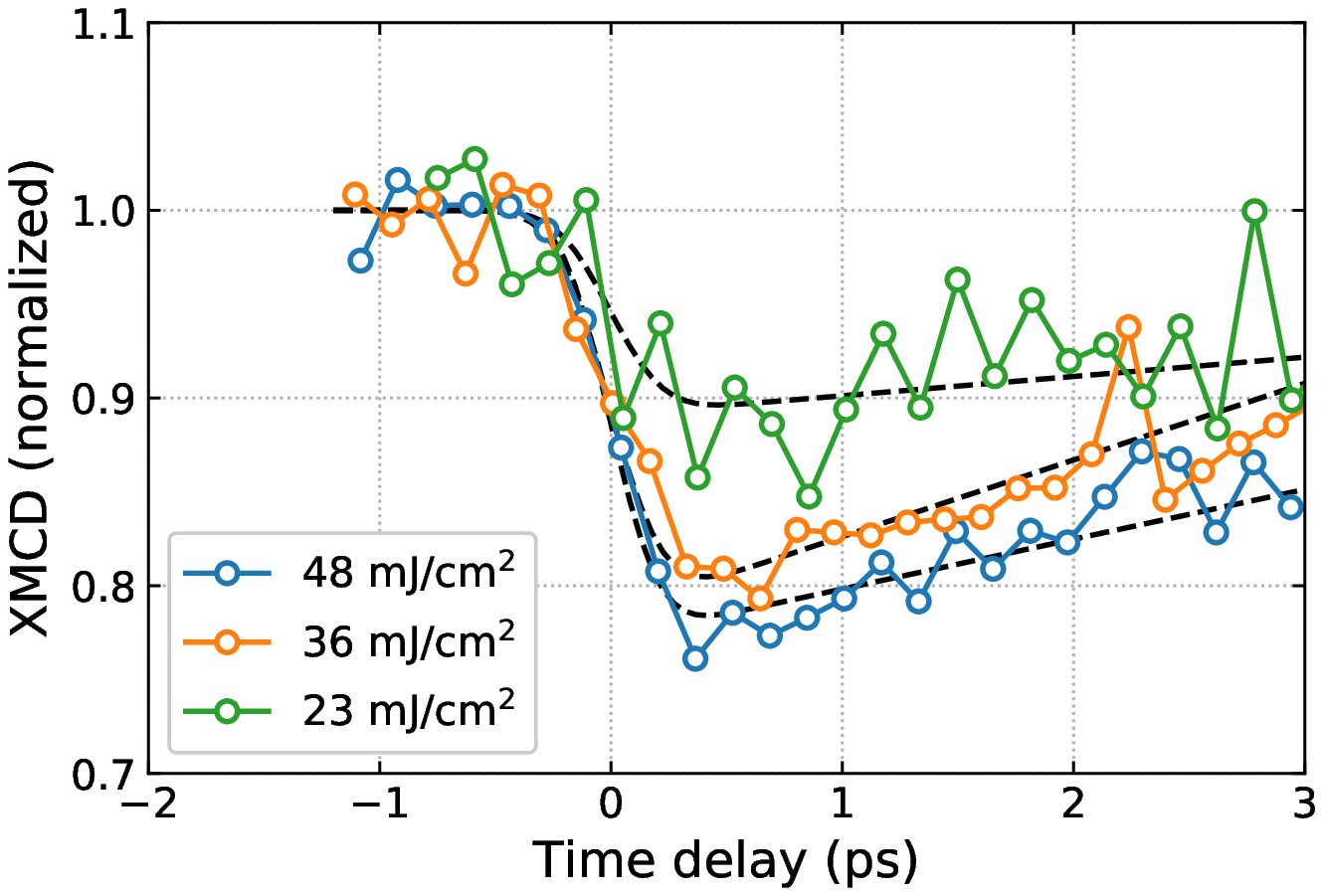}  
	\end{tabular}
	\caption{XMCD time scans on Ni/Ru/Fe with photon energy set to the Ni $L_3$ (left) and Fe $L_3$ (right) edges, for indicated pump laser fluences. The magnetic layers were oriented parallel through a magnetic field.}
	\label{NiRuFe-parallel-timescans}
\end{figure}

XMCD time scans on the Ni/Ru/Fe sample with parallel alignment of the Ni and Fe magentization are plotted in Fig.~\ref{NiRuFe-parallel-timescans}, layer resolved to display the (upper) Ni and (lower) Fe layer.

\section{Additional spectra on N\lowercase{i}/R\lowercase{u}/F\lowercase{e}}

\begin{figure}[htb!]
	\centering
	\begin{tabular}{@{}cccc@{}}
		\includegraphics[height=.56\textwidth]{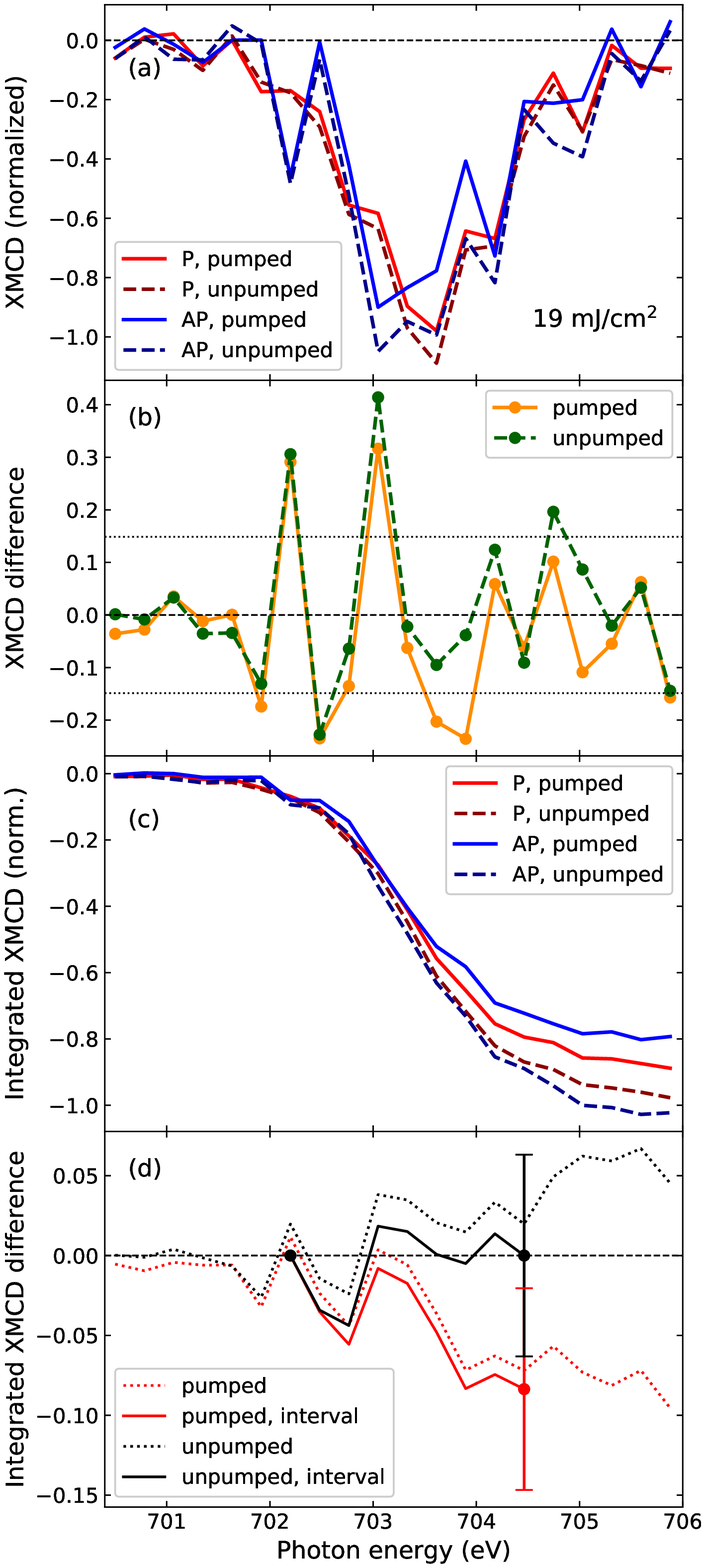} &
		\includegraphics[height=.56\textwidth]{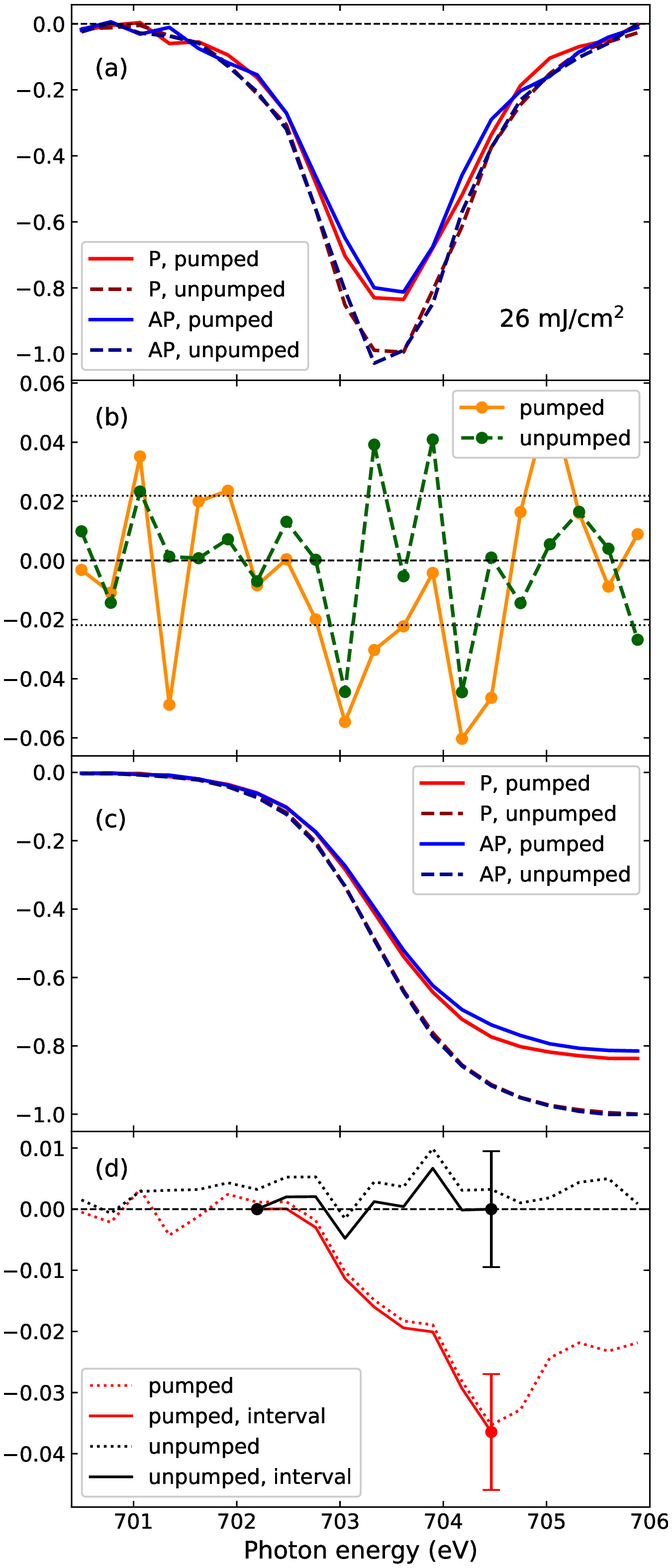} &
		\includegraphics[height=.56\textwidth]{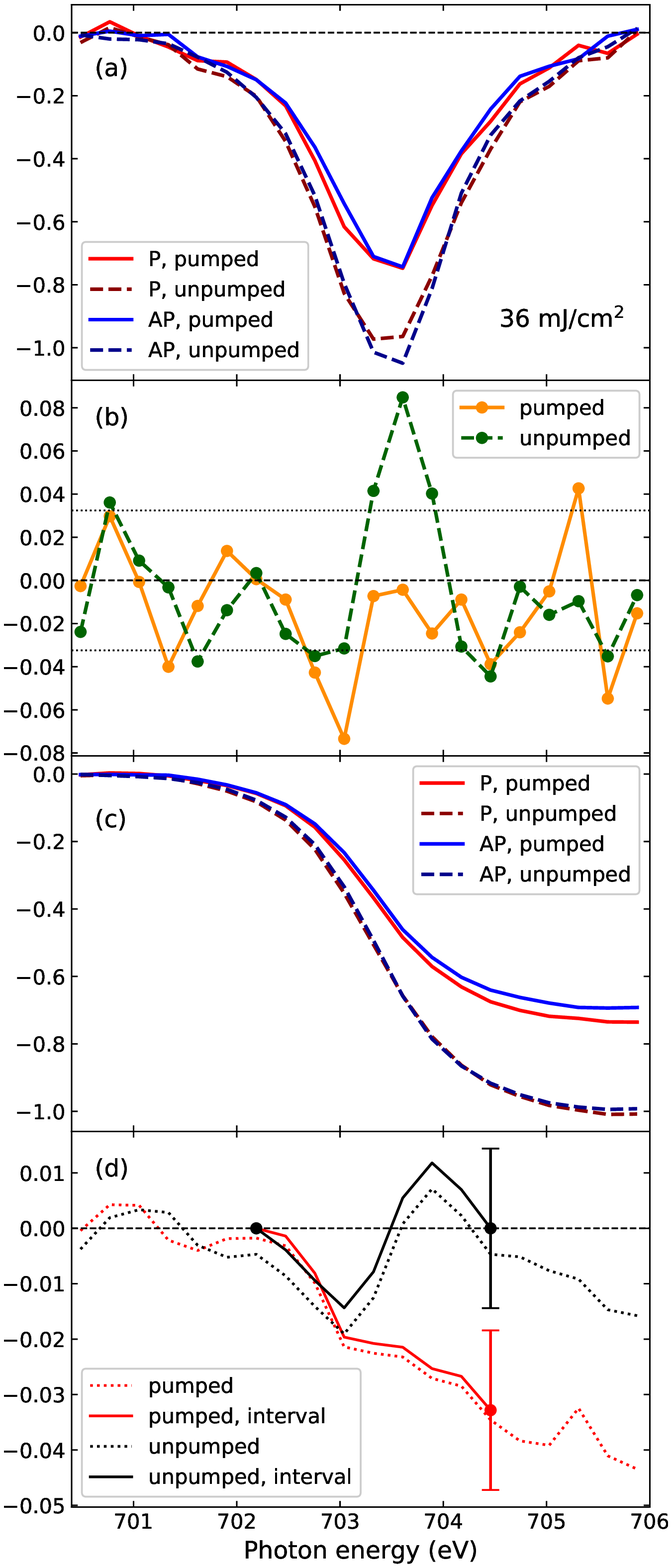} &
		\includegraphics[height=.56\textwidth]{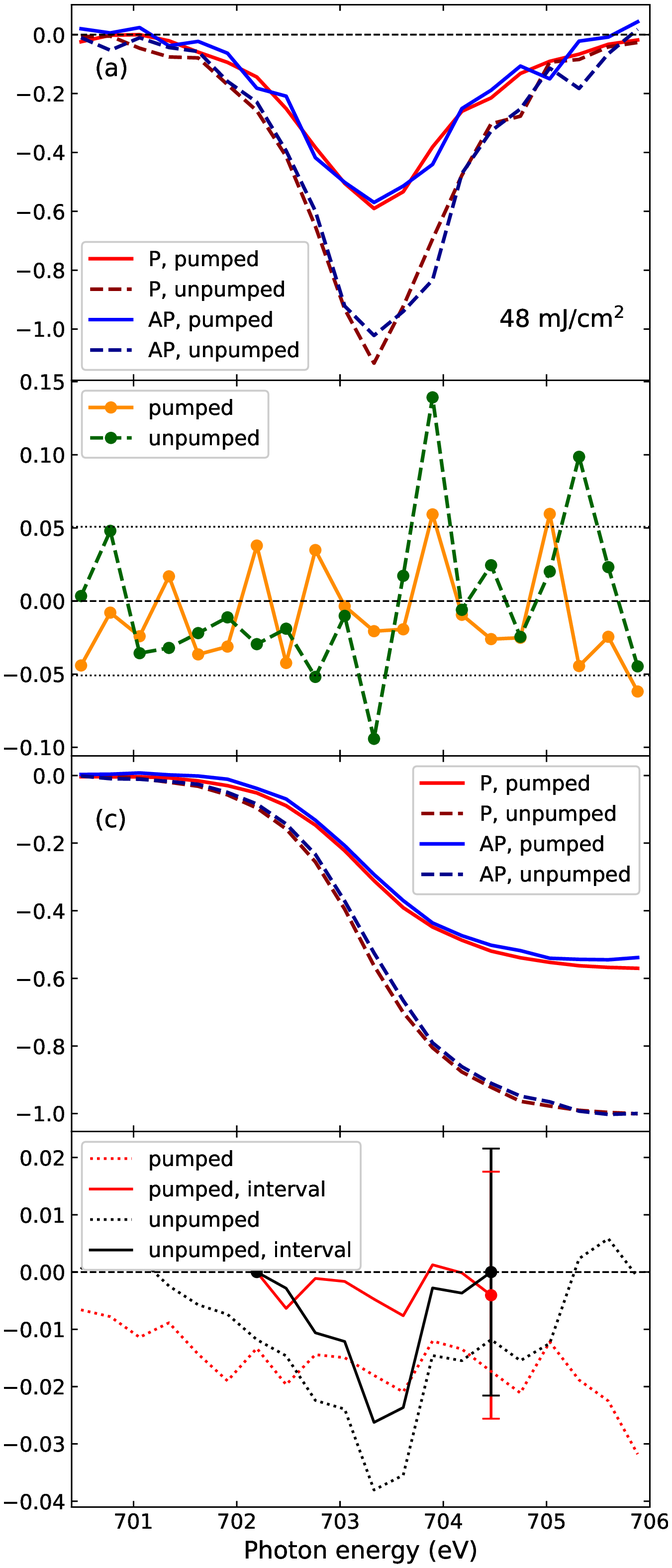} 
	\end{tabular}
	\caption{Ni/Ru/Fe XMCD measured at $\approx$1~ps after excitation. Four measurements with increasing pump fluence are shown, as indicated in the top panels. (a) XMCD of both relative magnetization orientations, parallel (P) and antiparallel (AP), for the pumped and unpumped sample. (b) XMCD difference of P and AP, (c) integration of the XMCD curves shown in (a), normalized to the value of the unpumped sample. (d) difference of the integrated XMCD. 
		Note the large XMCD fluctuations in the 19~mJ/cm$^2$ panel, causing a poor match of the unpumped curves in (c) and increased error margins in (d) for this fluence.}
	\label{NiRuFe-energy-scans}
\end{figure}

Besides time-scans recorded at fixed photon energy, we recorded XAS spectra around the $L_3$ absorption edges, applying four different magnetic fields. This forced the magnetization directions of the Ni and Fe layers into each of the four possible configurations, with two parallel and two antiparallel configurations. From these spectra, we calculate the dichroism for parallel and antiparallel configuration.
In order to obtain a common normalization of the two XMCD spectra, we also recorded for each spectrum  the sample in equilibrium, e.g.\ without pump laser pulse. The measurements for four different pump laser fluences are plotted in Fig.~\ref{NiRuFe-energy-scans}.

\section{\lowercase{fs} laser induced change in XAS}

\begin{figure}[!htb]
	\begin{tabular}{@{}cc@{}}
		\includegraphics[width=6cm]{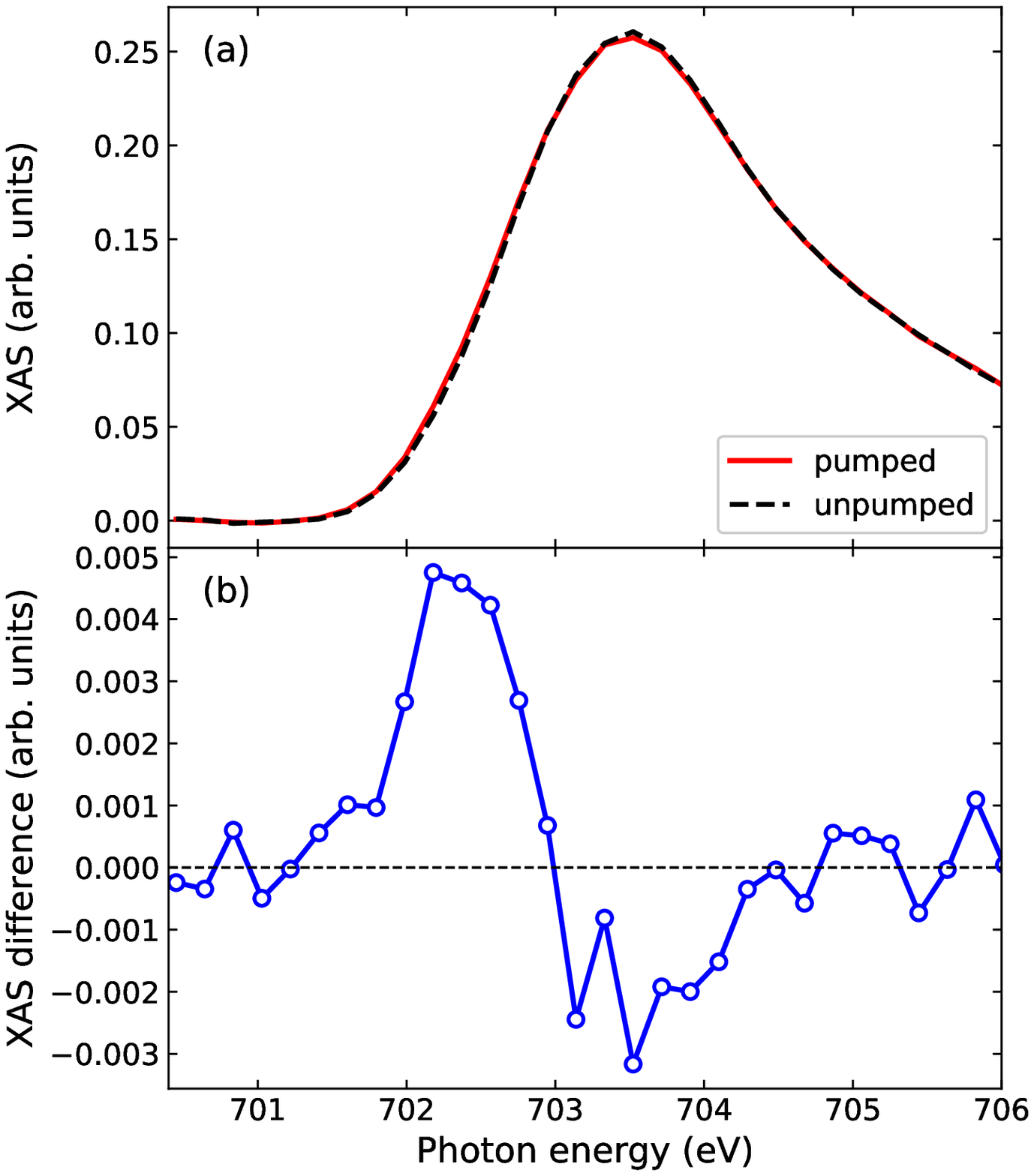} &
		\includegraphics[width=6cm]{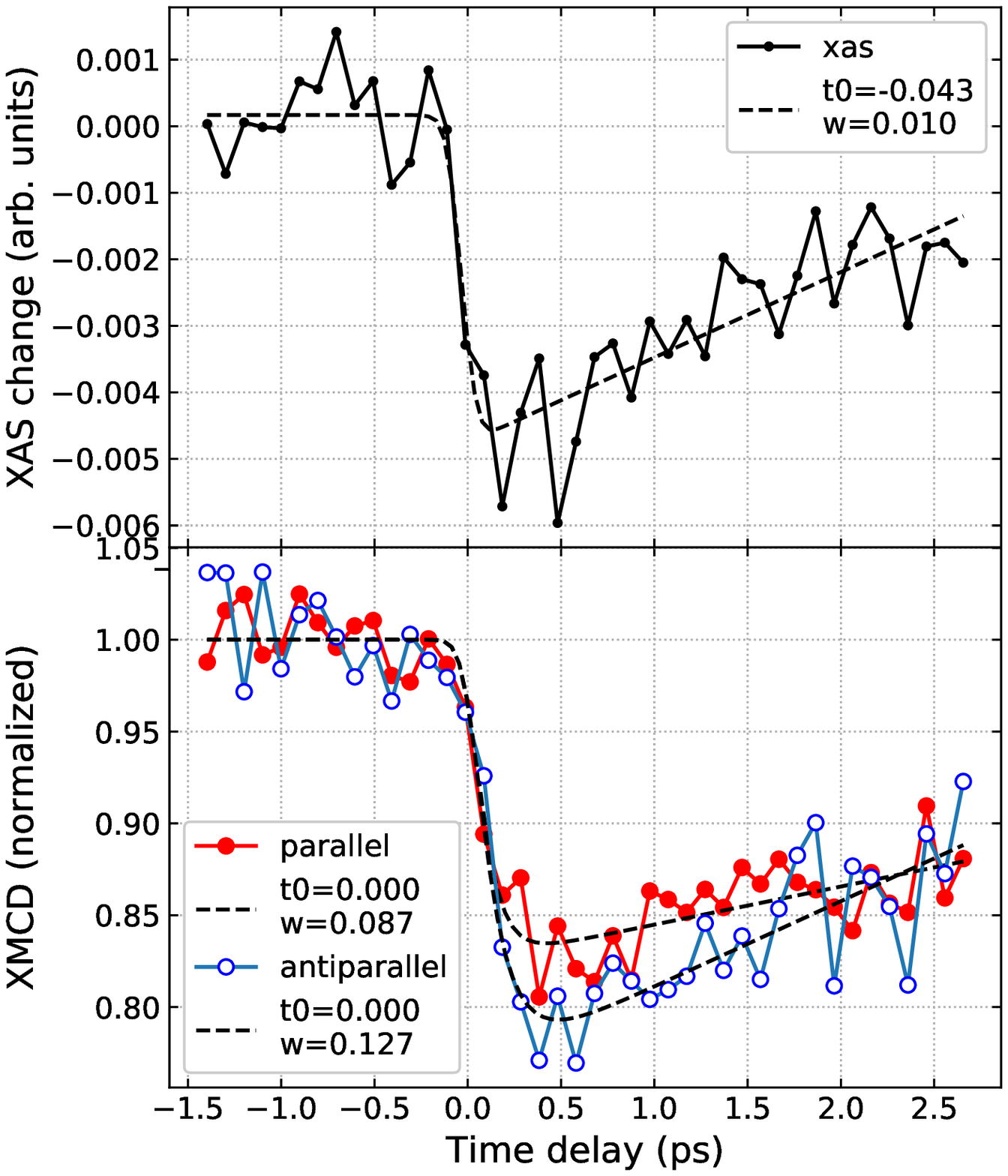}  
	\end{tabular}
	\caption{
		Left: 
		Ni/Ru/Fe XAS spectra at the Fe $L_3$ absorption edge: (a) laser pumped and unpumped absorption and (b) their difference.
		Right:
		Ni/Ru/Fe time dependent XAS changes and XMCD for parallel and antiparallel magnetization orientation, all measured at the Fe $L_3$ absorption edge, with 26~mJ/cm$^2$ pumping fluence.
	}
	\label{NiRuFe-XAS}
\end{figure}

Besides the obvious time resolved changes in the XMCD signals as presented in the main text, we observed subtle changes in the absorption lines, which may be seen as footprint of the laser excited electronic system. Similar as in previous fs XAS measurements on $3d$ transition metals \cite{Stamm2007,Boeglin2010}, a slight shift of the absorption line towards lower photon energies can be seen in Fig.~\ref{NiRuFe-XAS}(left).
The dynamic response of the XAS change directly relates to the XMCD dynamics as shown on the right in Fig.~\ref{NiRuFe-XAS}. These data were corrected for their timing jitter using the time tool data (last option in \ref{jitter-corr}), resulting in an improved time resolution at cost of a slightly increased intensity noise.

\section{Additional data on N\lowercase{i}/C\lowercase{u}/F\lowercase{e}}

\begin{figure}[!htb]
	\centering
	\includegraphics[scale=0.5]{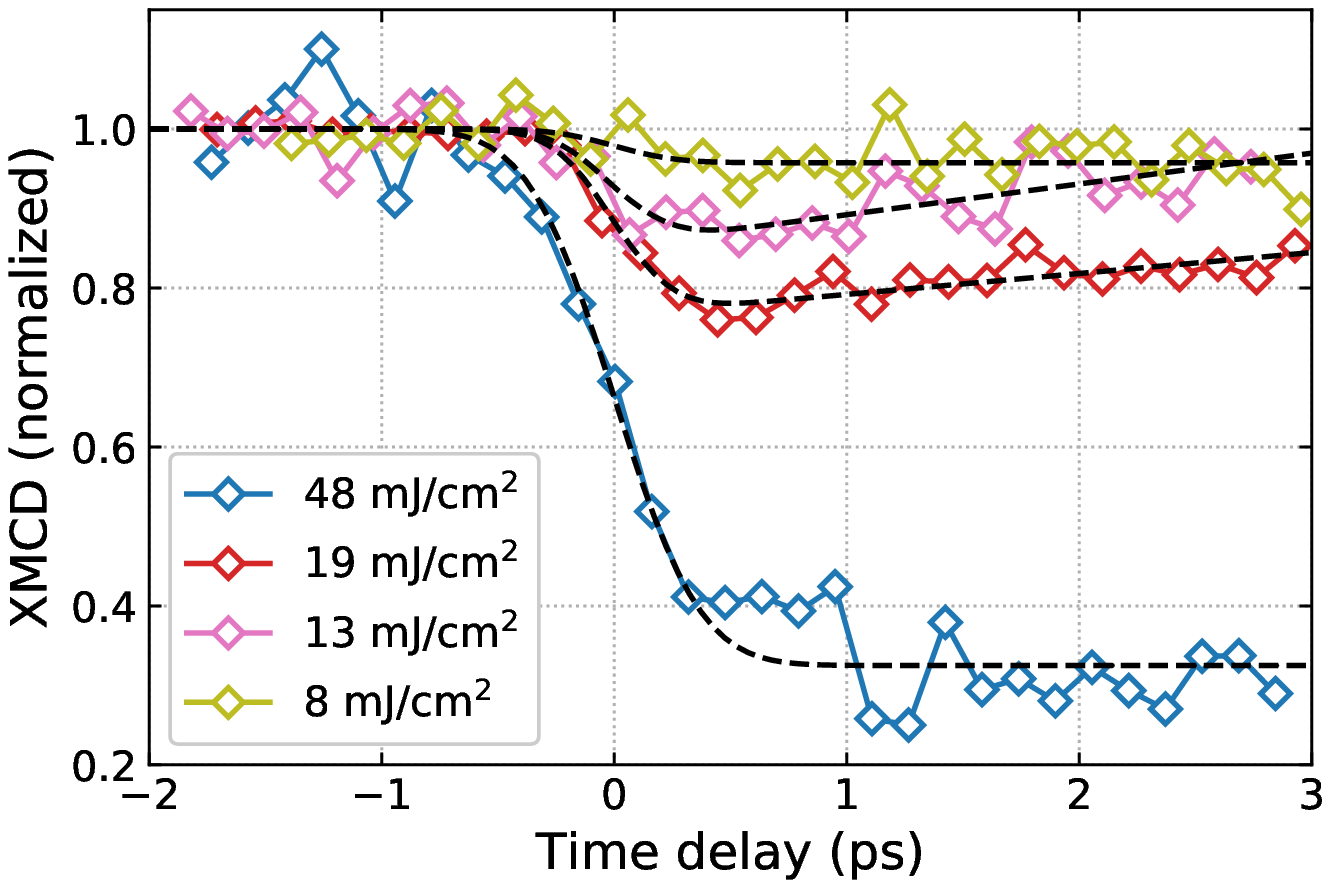} 
	\caption{Ni/Cu/Fe XMCD at the Fe $L_3$ absorption edge as function of pump-probe time delay, excited with the indicated pump laser fluence.}
	\label{NiCuFe-timescan}
\end{figure}

\begin{figure}[!htb]
	\centering
	\begin{tabular}{@{}ccc@{}}
		\includegraphics[height=.24\textwidth]{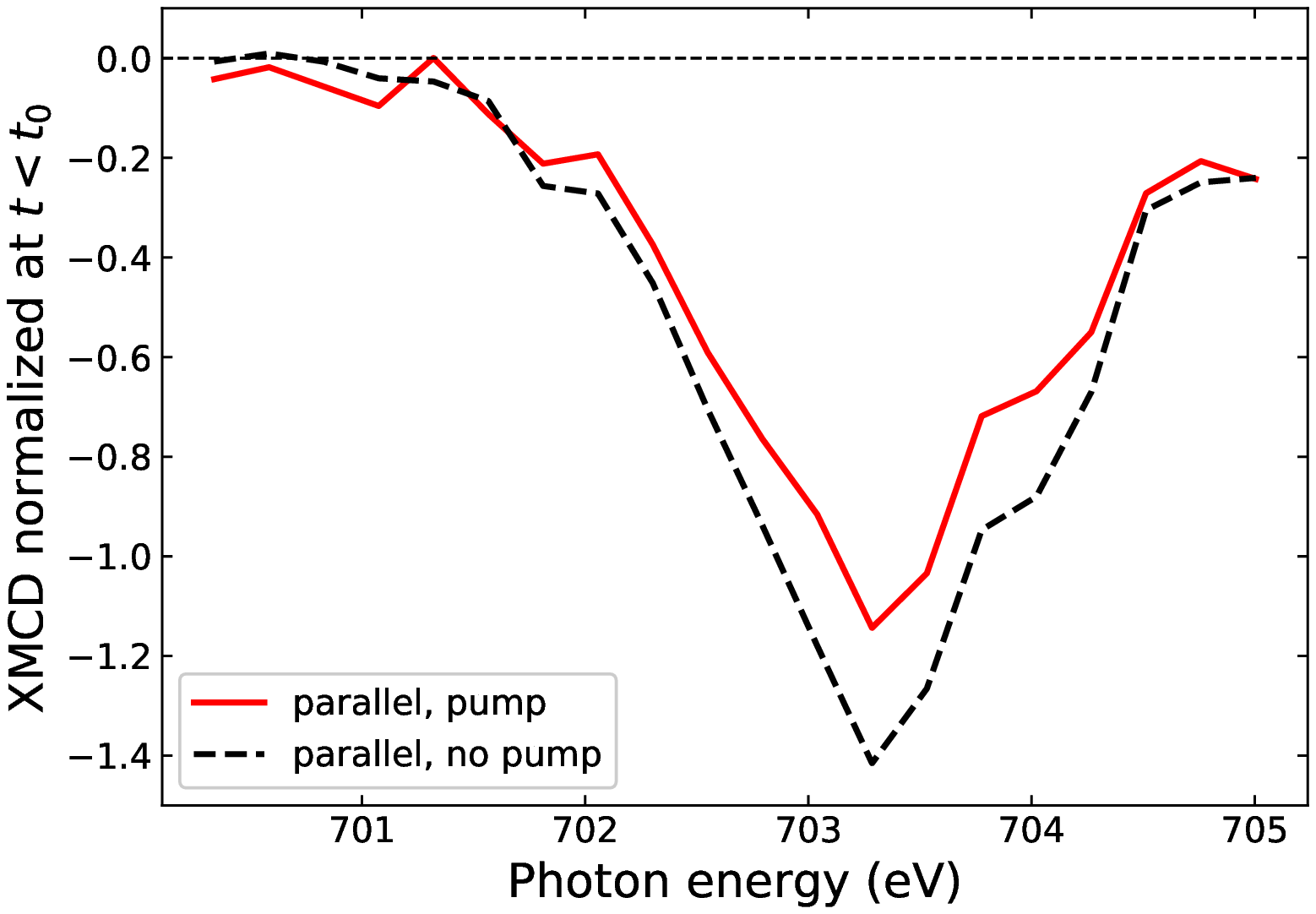} &
		\includegraphics[height=.24\textwidth]{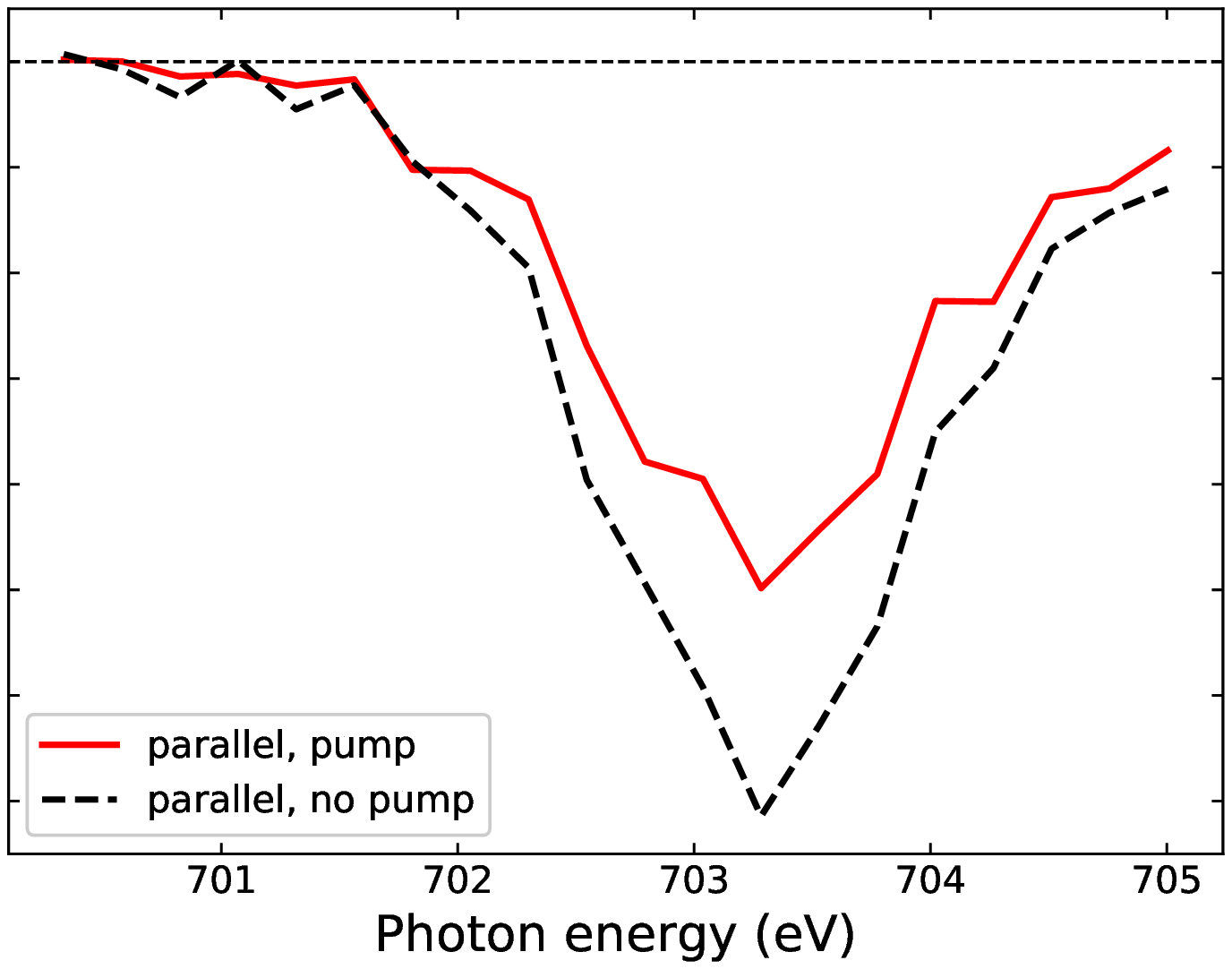} &
		\includegraphics[height=.24\textwidth]{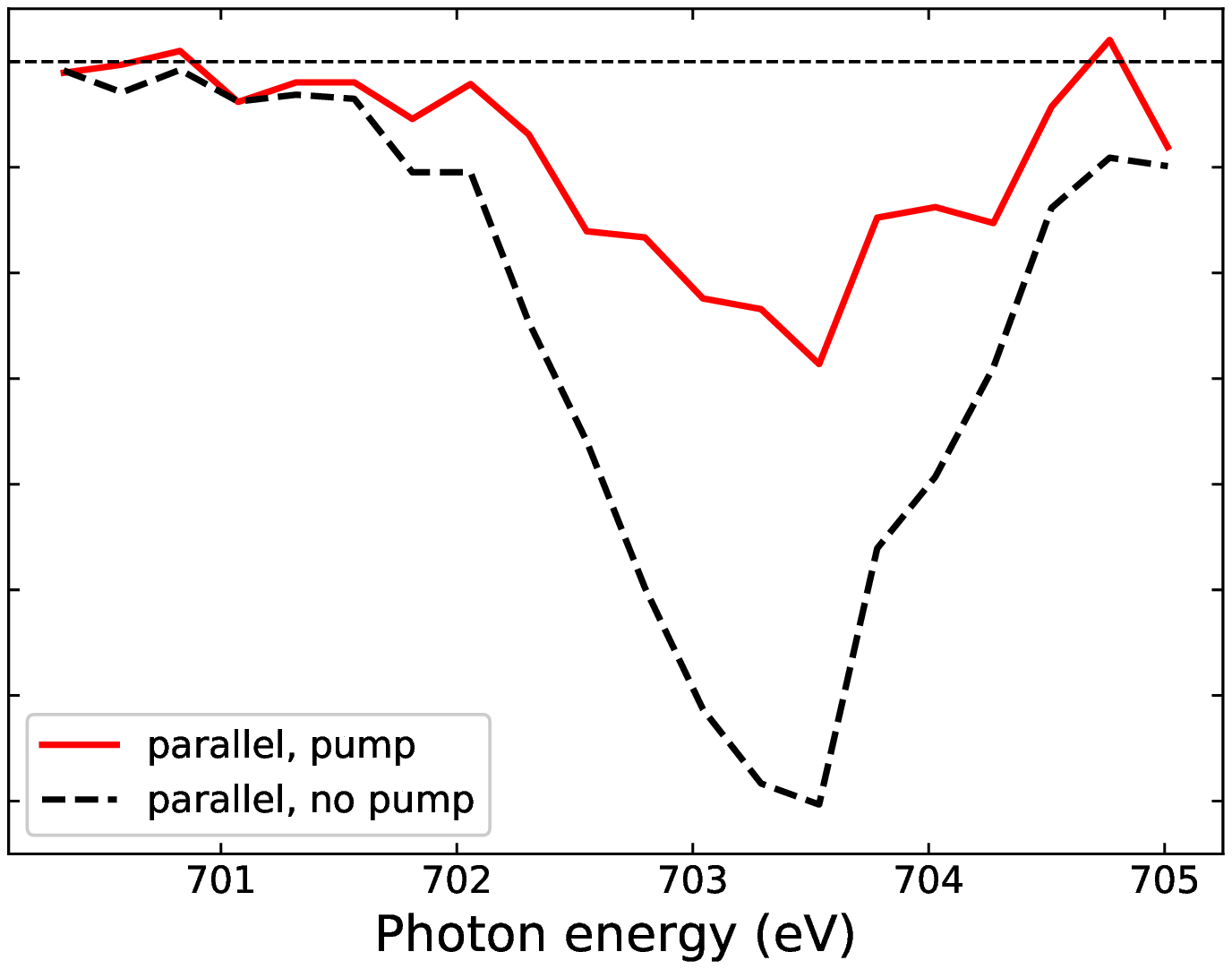} 
	\end{tabular}
	\caption{Ni/Cu/Fe XMCD at the Fe $L_3$ absorption edge, excited with the indicated pump laser fluence, (below) spectra measured at fixed time delay of $\approx$1~ps and at increasing fluences of 19, 26, and 48 mJ/cm$^2$ (left to right).}
	\label{NiCuFe-energyscan}
\end{figure}

Time resolved XMCD data on the Ni(5~nm)/Cu(30nm)/Fe(4~nm) is presented in Fig.~\ref{NiCuFe-timescan}, for different pump laser fluences. The Ni and Fe magnetization directions were always aligned parallel. XMCD spectra of the laser pumped and unpumped sample are plotted in Fig.~\ref{NiCuFe-energyscan} for increasing pump laser fluences.

\section{Modeling of the laser absorption in multilayer samples}

\begin{figure}[htb]
	\centering
	\begin{tabular}{@{}cc@{}}
		\includegraphics[height=.26\textwidth]{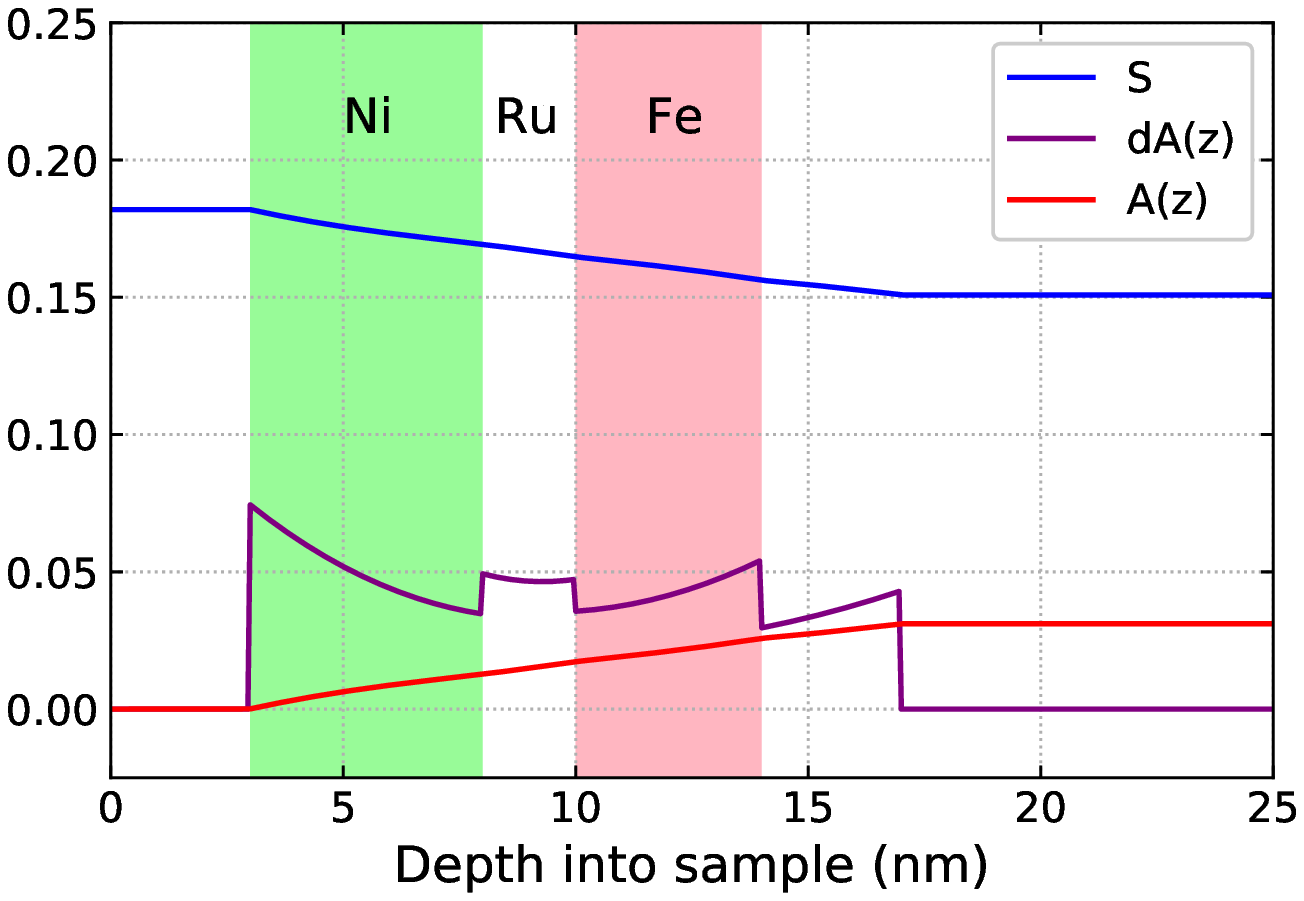} &
		\includegraphics[height=.26\textwidth]{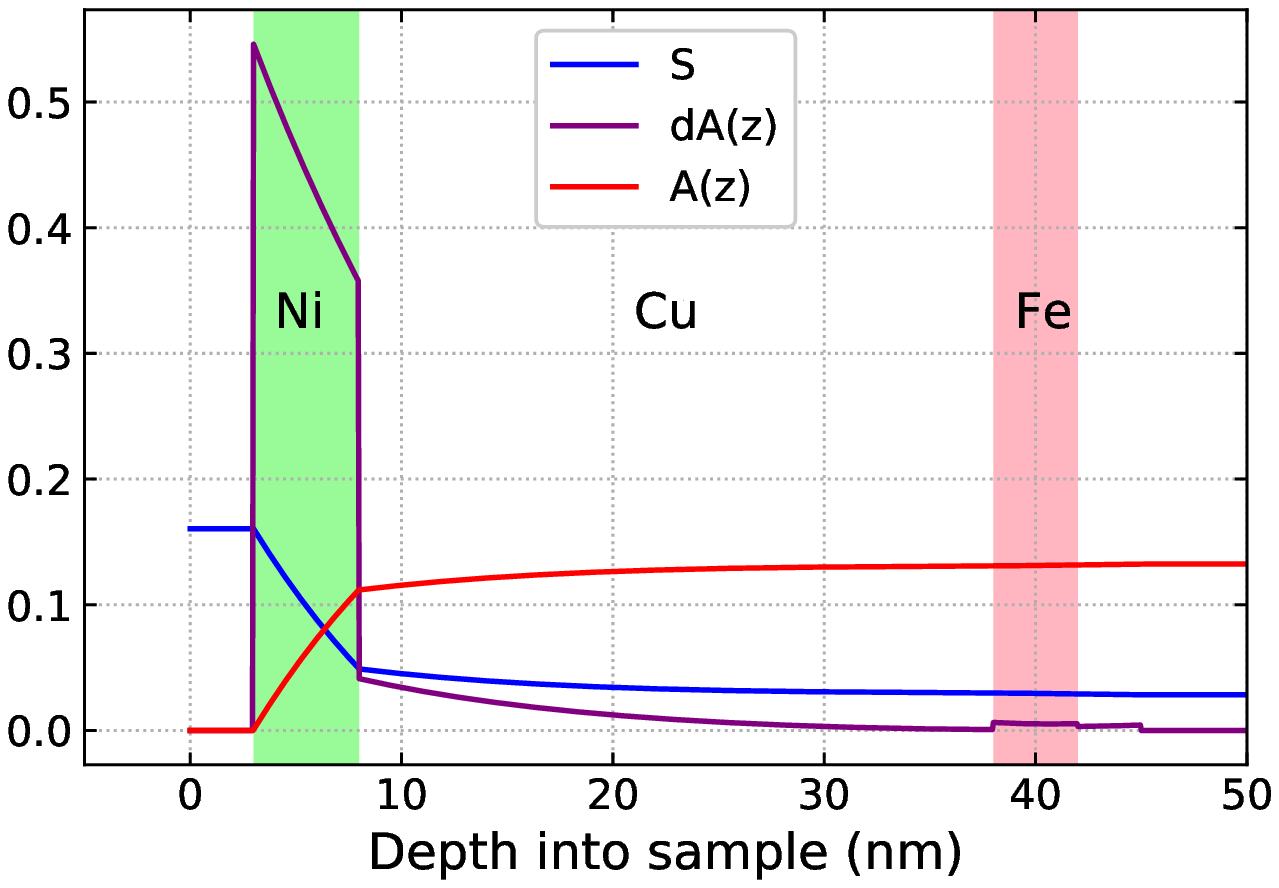} 
	\end{tabular}
	\caption{Simulated optical absorption in multilayered samples. Left, AlO$_x$(3)/Ni(5)/Ru(2)/Fe(4)/Ta(3)/Si$_3$N$_4$(188)/Al(100) and right, AlO$_x$(3)/Ni(5)/Cu(30)/Fe(4)/Ta(3)/Si$_3$N$_4$(188)/Al(100), with layer thickness in nm stated in parentheses.}
	\label{optical-absorption}
\end{figure}

The absorption of the exciting laser pulse is modeled within a transfer matrix calculation \cite{Byrnes2016}, in order to obtain the absorbed energy in each layer. Fig.~\ref{optical-absorption} shows the resulting Poynting vector $S$, the differential absorption $dA(z)$ and the depth-dependent absorption $A(z)$. The results are summarized in Table~\ref{absorption-table}. It becomes apparent, that for Ni/Ru/Fe the Fe layer is still affected by the pump laser beam, although not as strong as the Ni layer. In Ni/Cu/Fe, the absorption in Fe may be neglected. Both samples have a very strong overall reflectivity, which is due to the Al deposition on the back of the Si$_3$N$_4$ membrane. Especially the absorption in the magnetic layers in the Ni/Ru/Fe sample is strongly reduced. This is the reason for the rather high incident fluence for the laser excitation applied during the measurements. As a comparison, the last column in Table~\ref{absorption-table} states calculated values for a Ni/Ru/Fe sample that would have no Al layer on the back of the Si$_3$N$_4$ membrane. While the absorption in the magnetic layers is much stronger, the ratio $A(\mathrm{Ni})/A(\mathrm{Fe})$ is almost the same as for the measured sample with heat sink. Adjusting the incident pump laser fluence to account for the sample reflectivity, we thus get comparable excitation profiles across the Ni/Ru/Fe stack, with and without Al back coating.

\vspace{0.1 cm}
\begin{table}[htb]
	\setlength{\tabcolsep}{12pt}
	\begin{tabular}{@{}l l l l l @{}}
		\toprule[0.05em]
		Quantity &  & Ni/Ru/Fe & Ni/Cu/Fe & Ni/Ru/Fe (no Al)\\
		\midrule[0.025em]
		Sample absorption & $A$ & 0.1818 & 0.1608 & 0.4682 \\
		Absorption in Ni & $A(\mathrm{Ni})$ & 0.0125 & 0.1112 & 0.1860 \\
		%
		%
		Absorption in Fe & $A(\mathrm{Fe})$ & 0.0085 & 0.0011 & 0.1220 \\
		%
		%
		Absorption ratio & $A(\mathrm{Ni}) / A(\mathrm{Fe})$ & 1.47 & 101.3 & 1.52 \\
		Reflected intensity & $R$ & 0.818 & 0.839 & 0.2563\\
		\bottomrule[0.05em]
	\end{tabular}
	\caption{Simulated absorption of the pump pulse in the Ni and Fe layers. Last column: ficticious sample without Al layer.}
	\label{absorption-table}
\end{table}


%

\end{document}